\documentclass[runningheads]{llncs}
\usepackage[utf8x]{inputenc}
\usepackage{epsfig,amsmath}

\newcommand{\killonq}{Kilander and L\"{o}nnqvist}
\begin{document}

\pagestyle{headings}

\mainmatter

\title{Lemma 4:  Haptic Input + Auditory Display = Musical Instrument?}

\titlerunning{Lemma 4: Haptic Input + Auditory Display = Musical Instrument?}

\author{Paul Vickers}
\authorrunning{Paul Vickers}

\institute{Northumbria University,\\
School of Computing, Engineering, and Information Sciences,\\
Pandon Building, Camden St.,\\
Newcastle upon Tyne NE2 1XE, UK\\
\email{paul.vickers@unn.ac.uk}}

\maketitle

\begin{abstract}
In this paper we look at some of the design issues that affect the success of multimodal displays that combine acoustic and haptic modalities. First, issues affecting successful sonification design are explored and suggestions are made about how the language of electroacoustic music can assist. Next, haptic interaction is introduced in the light of this discussion, particularly focusing on the roles of gesture and mimesis. Finally, some observations are made regarding some of the issues that arise when the haptic and acoustic modalities are combined in the interface. This paper looks at examples of where auditory and haptic interaction have been successfully combined beyond the strict confines of the human-computer application interface (musical instruments in particular) and discusses lessons that may be drawn from these domains and applied to the world of multimodal human-computer interaction. The argument is made that combined haptic-auditory interaction schemes can be thought of as musical instruments and some of the possible ramifications of this are raised.  
\end{abstract}
\section{Multimodal Challenges}
Little research has been conducted into how haptic and auditory modalities can best be combined in human-computer interaction. However, in the non-computing world the two modalities have been partners for a long time. First Hollywood gave us \emph{movies} (the visual display in computing terms). Then came the \emph{talkies} (and auditory display in computing). In  \emph{Brave New World} (1932) Aldous Huxley offered us the notion of the \emph{feelies} -- the arm rests in theatre seats would provide haptic stimulation during erotic features. Whilst the feelies are not with us yet haptic displays are making inroads into the development of human-computer interfaces.  As the auditory display community has discovered there are several hurdles that must be jumped before a new interaction modality is considered acceptable. Haptic interaction also raises its own usability issues which are being dealt with by researchers. However, the combined use of auditory display and haptic input and output in a single application raises a new set of design challenges. The field is, perhaps, in an analogous position to that of cinema in the late 1920s. Early film sound was largely causal  and was thus a recording of the sonic events in the scene. The use of sound as a separate non-synchronous entity was advocated by Russian filmmaker Sergei Eisenstein whose 1928 \textit{Statement on Sound} \cite{Eisenstein:1988} suggested using sound as a counterpoint to the visuals. Eisenstein wanted to make montages of sound just as he used visual montage to great effect in his earlier silent films. The French filmmaker Ren\'{e} Clair was an early promoter of non-synchronous sound which has today developed into the disciplines of voice-overs/narration, sound effects \& Foley art, and film music. These early pioneers helped to make film a truly bimodal communication channel. Today, we are now beginning to move away from direct causal haptics and sonification in which the touch and auditory modes are used to accentuate visual information and are seeing new true multi-modal interfaces in which the different senses are used for separate but complementary information streams.

\subsection{Design Issues in Auditory Display}
The larger questions of sonification design are concerned with issues of intrusiveness, distraction, listener fatigue, annoyance, display resolution and precision, comprehensibility of the sonification, and, perhaps binding all these together, sonification æsthetics.

There is a tension in auditory displays between the sonification being perceptible to its intended audience and being too intrusive or annoying. In their work on awareness support systems, Hudson and Smith \cite{Hudson:1996} articulated the problem of intrusion in terms of awareness and privacy. They stated that this ``\emph{dual tradeoff is between privacy and awareness, and between awareness and disturbance}''. The more information an auditory display provides the richer the sonification yet the greater the potential for disturbance, annoyance, and an upset in the balance of the acoustic ecology. Gutwin and Greenberg \cite{Gutwin:1995} claimed sonification is a tradeoff between being well informed and being distracted. \killonq \ noted the effect of such sonifications on people sharing the workspace:``\emph{In a shared environment, one recipient may listen with interest while others find themselves exposed to an incomprehensible noise}'' \cite{Kilander:2002}. Indeed, commenting upon the design of their \emph{nomadic radio} system\footnote{In  \emph{nomadic radio} a mixture of ambient sound, recorded voice cues, and summaries of email and text messages is used to help mobile workers keep track of information and communication services.}, Sawhney and Schmandt \cite{Sawhney:1999,Sawhney:2000} cautioned that care must be taken to ensure that the auditory display intrudes minimally on the user's social and physical environment.

In dealing with intrusiveness, Pedersen and Sokoler \cite{Pedersen:1997} framed the problem as a balance between putting a low demand on attention versus conveying sufficient information. They studied this problem through an `\emph{ecology of awareness}' thus acknowledging the importance of the acoustic ecology of a sonification. Pedersen and Sokola made the  auditory, visual, and haptic representations of their {\sc aroma} system highly abstract -- abstraction would allow useful information to be communicated without divulging too many details that would violate privacy\footnote{The system communicated information about elderly householders to relatives in remote locations.}. It was hoped that abstract representations would be better at providing ``\emph{peripheral non-attention demanding awareness}'' \cite[p. 53]{Pedersen:1997}. It was also noted that such abstract representations lend themselves to being remapped to other media (what Somers \cite{Somers:1998a} would call \emph{semiotic transformation} -- similar to Eisenstein's \emph{transference}), or, in turn foster the accommodation of user preferences (an important aspect of \emph{{\ae}sthetic computing} -- see Fishwick \cite{Fishwick:2006}). Unfortunately, user studies showed that the abstraction led to users interpreting the representations in varied ways that were not always correct \cite{Pedersen:1998}.  Furthermore, \killonq\  \cite{Kilander:2002} warned that the ``\emph{monitoring of mechanical activities such as network or server performance easily runs the risk of being monotonous}'' a finding observed by Pedersen and Sokola who reported that they soon grew tired of the highly abstract representations used in {\sc aroma}. It is interesting that some of the blame was attributed to an impoverished {\ae}sthetic, the feeling being that involving expertise from the appropriate artistic communities would improve this aspect of the work. 

Cohen \cite{Cohen:1994a} identified a general objection to using audio: people in shared office environments do not want more noise to distract them. Buxton \cite{Buxton:1989} argued that as audio is ubiquitous it would be less annoying if people had more control over it in their environments. Lessons from acoustic ecology would be helpful here.
\subsection{Acoustic Ecology}
The term \emph{acoustic ecology} \cite{Wrightson:2000} comes from work begun by R. Murray Schafer in the 1960s as part of his \emph{World Soundscape Project} at Simon Fraser University (see \cite{Schafer:1977}). Schafer sees the world around us as containing ecologies of sounds. Each soundscape possesses its own ecology, and sounds from outside the soundscape are noticeable as not belonging to the ecology. In Schafer's worldview we are exhorted to treat the environments in which we find ourselves as musical compositions. By this we are transformed from being mere hearers of sound into active and analytic listeners -- exactly the characteristic needed to benefit most from an auditory display. When the environment produces noises that result from data and events in the environment (or some system of interest) we are able to monitor by listening rather than just viewing. In regard to auditory display the term acoustic ecology means the internal ecology of the various sounds within the sonification. That is, we treat the sonification both as a real-world soundscape in its own right, the acoustic ecology of which is jumbled, and as part of the wider real-world soundscape in which it is situated. Again, its sonic components may sit uneasily within the acoustic ecology of the host soundscape.

Cohen \cite{Cohen:1994a} defined an acoustic ecology as ``\emph{a seamless and information-rich, yet unobtrusive, audio environment}''. \killonq\  \cite{Kilander:2002} tackled this problem in their {\sc fuseONE} and {\sc fuseTWO} environments with the notion of a \emph{weakly intrusive ambient soundscape}, or {\sc wisp}. In this approach the sound cues for environmental and process data are subtle and minimally-intrusive\footnote{\killonq\  actually used the adjective `non-intrusive' to describe their sonifications. One could argue that this term is misleading as any sonification needs to be intrusive to some extent in order to be heard. Their term `weakly intrusive' is more helpful and more accurate.}. Minimal- or weak-intrusion is achieved in \killonq's scheme by drawing upon the listener's expectation, anticipation, and perception; anticipated sounds, say \killonq, slip from our attention.  For example, a ticking clock would be readily perceived and attended to when its sound is introduced into the environment (assuming it is not masked by another sound). As the steady-state of the ticking continues and the listener expects or anticipates its presence the perceived importance drops and the sound fades from our attention \cite{Kilander:2002}). However, a change in the speed, timbre, or intensity of the clock tick would quickly bring it back to the attention of the listener. Intrusiveness can thus be kept to a minimum by using and modulating sounds that fit well with the acoustic ecology of the sonification's environment. The sonification is discriminable from other environmental sounds (either by deliberate attentiveness on the part of the listener, or by system changes to the sounds) yet is sufficiently subtle so as not to distract from other tasks that the listener (and others in the environment) may be carrying out. To increase the quality of the acoustic ecology further, {\killonq} used real-world sounds rather than synthesized noises and musical tones. They concluded that ``\emph{easily recognisable and natural sounds \ldots \emph{[stand]} \ldots the greatest chance of being accepted as a part of the environment. In particular, a continuous background murmur is probably more easily ignored than a singular sound, and it also continuously reassures the listener that it is operative}'' \cite{Kilander:2002}.

The audibility of sonifications is an important factor and is tightly coupled to the issue of intrusiveness. The comprehensibility of sonifications depends on many factors including the production quality of the sounds, the quality of the playback system, and cultural and metaphoric associations. Many data require metaphoric or analogic mappings for audio representation as they do not naturally possess their own sound. The choice of metaphor may determine how learnable and comprehensible the mapping is. For example, {\killonq} found that the sound of a golf ball dropping into a cup was difficult for listeners to recognize ``\emph{except possibly for avid golfers}'' \cite{Kilander:2002} whilst the sound of a car engine was easy to identify. This highlights the fact that when using real-world sounds it is important to assess the cultural attributes of those sounds. Investigating musical tones for the monitoring of background processes S{\o}r{\aa}sen \cite{Sorasen:2005} found that sudden onset or disappearance of a timbre is easier to detect than changes in the rhythm and melody of that timbre. He concluded: ``\emph{changes within one single instrument should be very carefully designed to represent non-binary changes in state or modus}''.

\section{Applying Musical Æsthetics}
Vickers and Hogg \cite{Vickers:2006a} argued that the æsthetics and composition approaches of electroacoustic and musique concrète\footnote{Musique concrète is a branch of electroacoustic music pioneered by Pierre Schaeffer (1910-1995) in which music is produced by editing together processed fragments of natural and industrial sounds.} may potentially lead to great success in sonification design given their dependence upon the gestural encoding present in sounds. According to Smalley's \cite{Smalley:1986} spectro-morphological\footnote{In musique concrète and electroacoustic music conventional pitched tones are only a subset of spectro-morphologies within a much broader world of spectra \cite{Smalley:1986}. The reliance on architectures based around harmonic progressions of pitches is removed.} classification we hear the physical, gestural qualities in sounds, and these in and of themselves (though usually in combination with timbre and volume) carry sufficient information regarding movement, atmosphere, size, material quality, and so forth to offer information to the perceiver that serves to generate meaning akin to that generated by musical harmonic/tonal systems. It has the added advantage of being arguably less culture specific, that is it is not classical, or pop, or anything we already recognize -- it is rather a system that is more open to reading than it is a musical style that is recognised as such. Criticisms of cultural imperialism are often raised when sonifications based upon tonal (or even atonal or Schoenbergian serialist) structures are presented at conferences. The potential offered by electroacoustic musical forms to avoid cultural stereotyping means much more serious consideration should be given to their use in auditory display research.

\begin{figure}[hbt] 
\includegraphics[width=\linewidth]{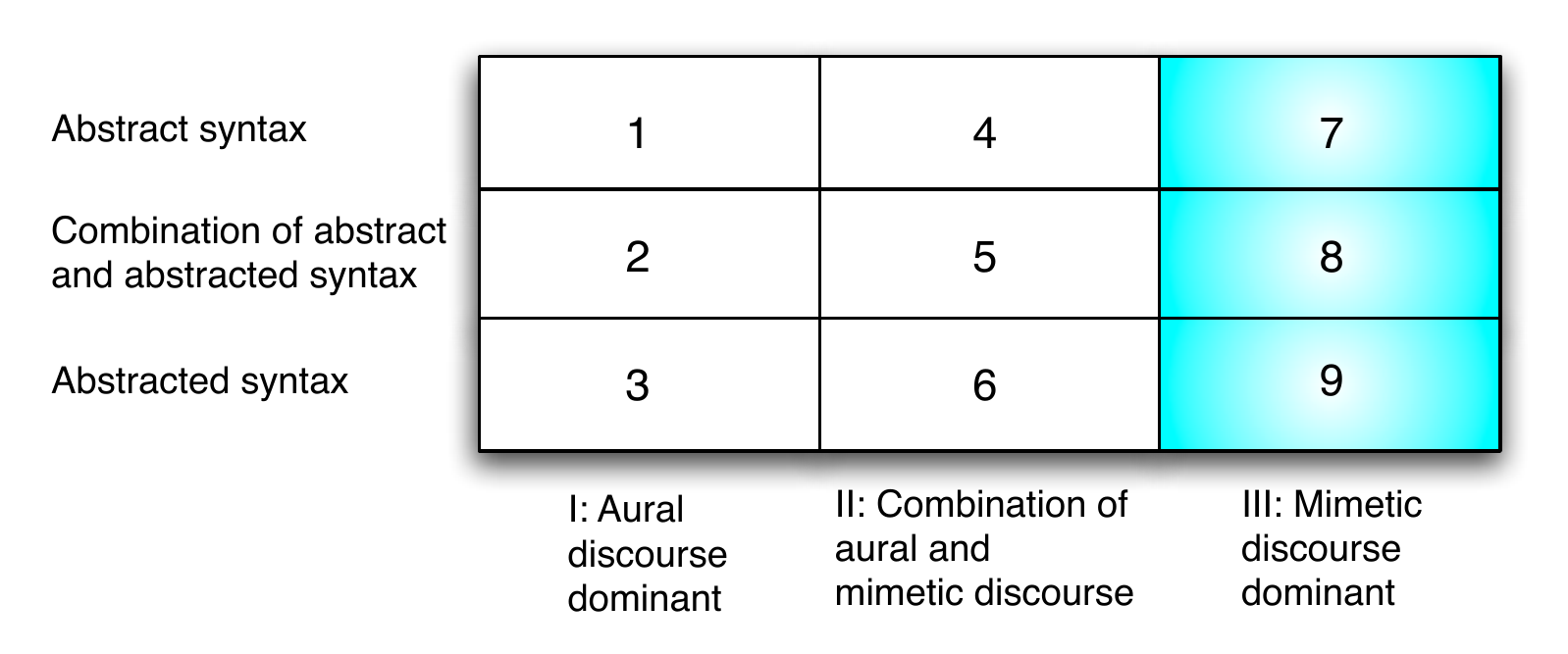} 
\caption{Emmerson's Language Grid \cite{Emmerson:1986}} 
\label{fig:grid} 
\end{figure}

To help make the link between auditory display and electroacoustic music clearer, let us consider Emmerson's \cite{Emmerson:1986} \emph{language grid}  in which he classifies electroacoustic music into a nine-sector space on two axes: the level of syntactical abstraction of the music and the use of mimetic reference vs. aural discourse (see Fig. \ref{fig:grid}). Mimetic sound imitates or represents nature and aspects of human culture.
In the language grid pitch-oriented music (both tonal and atonal) occupies grid sector 1 (abstract musical syntax with aural discourse dominant). We have shaded positions 7, 8, and 9 because they bound the region occupied by sonification. Sonifications are by definition mimetic in that their goal is to represent objects, events, or data of the domain in question. Some sonifications (sector 7) are grounded in the tonal music/melodic paradigm and so use \emph{abstract} syntax\footnote{In Emmerson's classification an \emph{abstract} syntax is one in which the musical ideas have been organised and constructed independently from the sound materials. An \emph{abstracted} organisation is one in which the music is abstracted from the sound-generating materials themselves. Thus, traditional music composition uses abstract syntax in which the notes on the score are related only to each other.}  (e.g. Vickers and Alty's {\sc caitlin} program sonification system \cite{Vickers:2005c} whilst others (sector 9) rely more on sounds \emph{abstracted} from real objects (e.g. auditory icons \cite{Gaver:1986}. In region 8 are those sonifications that use both abstract  and abstracted syntax (such as Barra et al.'s {\sc WebMelody} \cite{Barra:2002}). Sonification, then is concerned with mimetic discourse along the abstract-syntax/abstracted-syntax dimension. 
\subsection{Indexicality}
\begin{figure}[hbt] 
\centering
\includegraphics[width=\linewidth]{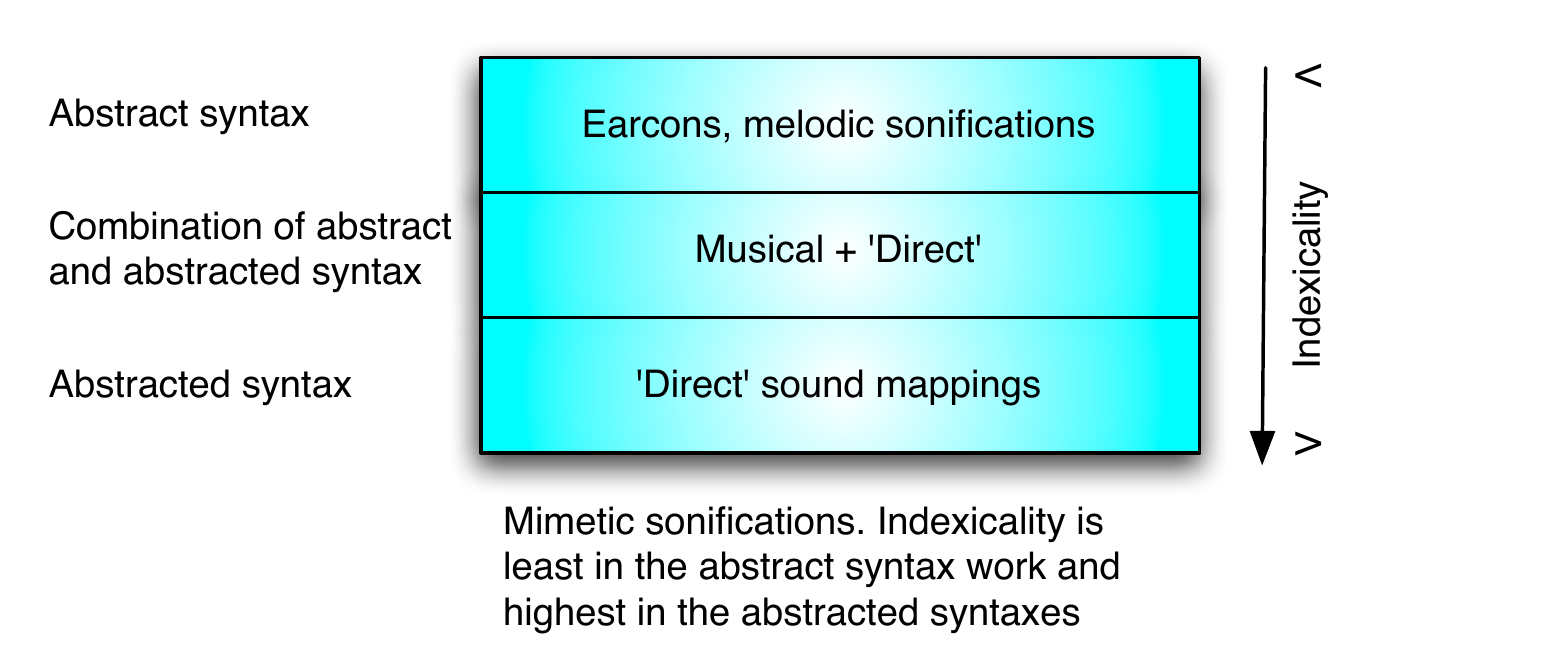} 
\caption{Sonification Indexicality (adapted from Emmerson's \emph{Language Grid} \cite{Emmerson:1986})} 
\label{fig:indexgrid} 
\end{figure}

Vickers and Hogg \cite{Vickers:2006a} introduced the idea of \emph{indexicality} to sonification discourse. Indexicality is associated with mimesis as it is a measure of how strongly a sound sounds like the thing that made it. In sonification practice indexicality is related to whether sonifications make more use of direct data-to-sound mappings (high indexicality -- the sound is derived directly from the data) or more use of metaphoric or interpretive mappings (low indexicality). Hayward's \cite{Hayward:1994} auditory seismograms are an example of the former in which seismographic data were scaled and frequency-shifted until they lay in the human audible range. Vickers and Alty's \cite{Vickers:2003} program sonifications are metaphoric: tonal musical motifs were used to stand for data and objects. From this, we can adapt Emmerson's language grid to the sonification domain to give Fig. \ref{fig:indexgrid} in which Emmerson's abstraction dimension has been retained but only the mimetic sectors have been carried over. In addition, the polarity of indexicality is indicated. Note that sonifications relying predominantly on abstract syntax (e.g. Vickers and Alty's program sonifications) possess lower indexicality than those making use of predominantly abstracted syntax (e.g. Hayward's seismograms).

There is a mapping, then, between sonifications and music compositions in that direct sonifications and concrete music possess high indexicality and metaphorical sonifications and abstract music possess low indexicality. That is, `direct' (abstracted syntax) and `metaphorical' (abstract syntax) in the sonification domain map to `concrete' and `abstract' respectively in the music domain. Musicologists would argue that music is as much a construct of the listener’s mind as of the composer's; if the listener perceives something as music then it \emph{is} music (though Smalley \cite{Smalley:1986} claims that the listener must  ``\textit{discover a perceptual affinity with its materials and structure}'' in order for this to happen). Thus, Vickers and Hogg \cite{Vickers:2006a} offer us:
\begin{lemma} $Sonification \implies Music$\end{lemma}
Kramer \cite{Kramer:1994b} notes the similarity in structure between sonification and music creation: sonification renders data in sound to allow a human listener to detect and comprehend patterns and structures in that data, whilst a musician renders a musical score so as to make it audible and thus make perceptible the music's structure and even give clues as to the composer's and the musician's emotional states. Thus, a piano is a sophisticated auditory display machine \cite{Vickers:2006a}: through the agency of the musician, the piano renders in sound (albeit in a highly complex and abstract way) the score, the technique of the musician, the physics of the piano, the emotional state of the musician and the composer, and even the musician's response to the feedback loop offered by his own ears. This gives us \cite{Vickers:2006a}:
\begin{lemma}$Music \implies Sonification$\end{lemma}
Thus, Vickers and Hogg \cite{Vickers:2006a} argue that sonification and music are mutually implicated and thus we get the logical biconditional:
\begin{lemma}$Sonification \iff Music$\end{lemma}
That is, if something is music then it is also a sonification and vice versa \cite{Vickers:2006a}. As an illustration of the link between music and sonification consider Barra et al's {\sc WebMelody} system \cite{Barra:2002}. Drawing on the ideas of futurist composer Luigi Russolo (1885-1947), principles of Pierre Schaeffer's musique concrète, and inspired by Edgard Varèse's \textit{Poème Électronique} (1958) and John Cage's aleatoric compositions (e.g. \textit{Music of Changes} (1951)), Barra et al \cite{Barra:2002} tried to construct sonifications for monitoring a web server that were ``\textit{neutral with respect to the usual and conventional musical themes.}'' They attempted to move away from the idioms of tonal and atonal (serialist) music and towards the more concrete compositions found in the musique concrète and electroacoustic traditions. 

\subsection{Acousmatics to Haptmatics}
One device employed very successfully by {\sc WebMelody} (and many other sonification systems) is \emph {acousmatic sound}. The term was introduced by Pythagoras who reputedly taught his students whilst standing behind a screen\footnote{Jérôme Peignot (1955) and François Bayle (1974) reintroduced the term in respect of musique concrète.}. Acousmatic sound, then, is that which one hears without the originating cause being visible to the listener. Auditory display research is replete with acousmatic sound as sonifications are often designed to highlight unseen or hidden data or events. We may also use haptic feedback to communictate information about unseen (or unheard) objects, data, or events (e.g. notification of incoming email). As this is an analogue of acousmatic sound we shall call it \emph{haptmatic} sensation. An example of an existing haptmatic display technique is Brewster and Brown's \cite{Brewster:2004} \emph{tacton} -- a haptic icon (c.f. auditory icons and earcons).

In fact, there is a very direct relationship between sonification and the kind of haptic feedback found in tactons and other similar approaches. Sound is simply a transverse wave with properties of frequency, phase, timbre, and amplitude. The vibro-tactile display offered by the tacton possesses similar properties -- indeed, if a tacton emitter is placed on a resonant surface it becomes a loud speaker. Sound and touch are very closely related in our everyday experience, especially in the region of low frequency sounds which are easily conducted through floors and other surfaces. At night clubs we can feel the bass frequencies in our chest; on the street we can feel heavy lorries approaching. The relationship between touch and sound is not only in this direction. In the next section we look at the relationship between haptic (gestural) input and sonification.
\section{Spectro-Morphology and Haptic Input as Performance}
Smalley's \cite{Smalley:1986} classification of sounds via spectro-morphology  provides us with an approach to sound and musical structures that focuses on the spectrum of available pitches and frequencies and how they are shaped (morphed) over time. In Smalley's model every sound contains gestural information that codes for the identity of movement, atmosphere, size, material quality, and so forth of the sound object in question. We are used to listening to sounds and decoding such gestural information. We can tell how fast a car is approaching us; we can estimate how hard a drum was hit and what size the drum is. Indeed, there is a very strong association between gesture and interpretation. This is especially true of music performance in which very complex and subtle gestural control is used to shape the music we hear and experience.

One of the challenges faced by audiences of electroacoustic music concerts is the lack of visible gestural interaction between the musician and the instrument. When the instrument is tape decks and laptop computers the familiar frame of reference of the direct relationship between a performer bowing strings, striking keys, moving sliders, and banging drums and the resultant sound is lost. It is replaced by a visible gesture set that often, at best, has a seemingly indirect relationship to the sonic experience and, at worst, no perceptual link at all (a reiteration of Eisenstein's contrapuntal sound, perhaps).

The `traditional' musical instrument has been with us so long that it has become the natural mechanism for making music and we are well attuned to watching musicians physically play their instruments. We note that for every input gesture there is a corresponding output sound. The expert listener and observer notes the subtlest of gestures but even the neophyte can observe the link between gross gestures (such as strumming a guitar) and the overall sonic output. 

How, then, does this inform us about the haptic control of interfaces that also use sonification? Spectro-morphology tells us that we listen for and can identify gestures in a sound. Experience shows us that we are used to associating physical gestures with consequent sound. If we are so sensitive to gesture, it might be that the haptic gestures become associated with the system's auditory output whether they are directly related or not. Might the subtle gestures needed for fine control of systems such as Sensable's Phantom\footnote{See \texttt{http://www.sensable.com/products/phantom\_ghost/phantom.asp}} be translated in the mind of the listener as being somehow related to any auditory display? We have shown above Vickers and Hogg's assertion that sonification and music are in a mutual implication relationship. If a sonified interface is thus a music playing system and the interface also provides gestural input in the form of haptic control, we suggest:
\begin{lemma}Auditory Display + Haptic Input = Musical Instrument\end{lemma}
That is, because of the ubiquitous frame of reference from musical instruments in which input gestures result in musical output we argue that a device/application interface that combines auditory display with haptic input might be viewed or interpreted as a musical instrument in the \emph{mind} of the user. It does not necessarily matter that the sonification produced by the system is not related either directly or indirectly to the haptic input (though a lack of even an indirect relationship is questionable) as one would still naturally seek links between the gestural inputs and the sonic outputs. Where there is no direct relationship it becomes quite possible that artificial cause-effect relationships will be constructed in the user's mind which may cause usability problems. A user moves a lever, grabs a virtual object, or makes a circular gesture to a mobile phone's camera, and the system happens to emit a sonified data stream immediately afterwards, it is very possible that a causal link will be established in the user's mind even if no link is present. 

\subsection{Haptic Input \& Auditory Display}
In other words, a musical instrument is a system in which manual gestures result (however visibly indirectly) in musical output. A multimodal system that combines haptic input with auditory display may thus be considered to be a musical instrument. If causal relationships between the haptic control and the auditory display are intended, then system designers might find it useful to draw lessons from the musical instrument design community for ways of improving the interaction\footnote{Where direct links do exist between the input and the output, there is also scope for users playing the interface like a musical instrument just for fun.}. If causal relationships between gestures and audio are not intended or desired (Eisenstein again), then interface designers also need to be aware of the dynamics of instrument design in order to try to `design out' any user perceptions of causality that arise through familiarity with the musical instrument paradigm. 
\subsection{Haptic Output \& Auditory Display}
Where systems combine auditory display with haptic output, other possible interactions between the two modalities arise. We commented above about the relationship between sound and touch, how certain sounds also generate vibrations that are perceived physically. Again, the question arises whether causal relationships between the haptic and auditpory outputs are inferred by the user.  Where no causal relationship exists such inference would likely be detrimental to the user's interpretation of the system's outputs. Again, the system might be considered a musical instrument. 
\section{Further Study}
Researchers who have reported the most success with their sonifications also tended to deal directly with the issue of the æsthetics and acoustic ecology of their sonifications. As the role of æsthetics is increasingly entering the consciousness of designers of computing systems (e.g., see Fishwick \cite{Fishwick:2006}) so it needs to inform the work of the auditory display community. It has been proposed that sonifications be viewed as works of musical art as they could then benefit from the application of the æsthetic practices employed by artists \cite{Vickers:2006a}\footnote{St. Augustine's Confessions act as a valuable cautionary tale against going too far down the art for art's sake route.}. When haptics are added to the mix there is a great potential for causal associations between the sound and the touch to be created in the user's mind. What is needed then, is research that explores the cognitive and artistic issues surrounding our haptic-sonification musical instruments. In what ways does haptic control affect the way we perceive an auditory display? If force feedback and sonification are used in tandem, does the brain automatically assert a link between the two data streams even when none is present? In the world of television and the movies, it is usually the case the the soundtrack affects the perceived meaning of the visual track and not the other way around. Will sound also have such an influence on our perception of force feedback?

It would be instructive to look at the semiotics of haptic-auditory interfaces. Whilst much has been written about the semiotics of visual and auditory messages less attention has been paid by semioticians to touch \cite{McGee:2003}. However, as musicologists and designers of new interfaces for musical expression\footnote{See \url{www.nime.org}.} are very interested in the relationships between physical gesture and sound with the machine as intermediary \cite{Magnusson:2006}, it may benefit the HCI community to temporarily cast their multimodal interfaces as musical instruments and see what design lessons can be learnt by studying these systems as a music researcher might. 
\emph{Speech Act Theory} \cite{Winograd:1986} studies how people use language (rather than establishing the truth-value of statements) \cite{McGee:2003} and offers three dimensions of act: \emph{locutory}, \emph{illocutory}, and \emph{perlocutory}. The locutory dimension deals with material aspects of an act's generation (e.g. strength of a physical movement). The illocutory aspects are to do with the intention behind an act. The percolutory dimension deals with the effect an act has upon the receiver.  By drawing together the skills of the HCI practitioner, the music researcher, and the semiotician, we may be better placed to understand the locutory, illocutory, and perlocutory nature of haptic and auditory signals and thus able to explore the rich interactions and the acousmatic and haptmatic effects that will result in the new generation of multimodal systems.
\bibliographystyle{splncs} 

\end{document}